\documentclass{iau_FM}
\usepackage{graphicx}
\usepackage{hyperref}

\title[] 
{An Astronomers Guide to Machine Learning}

\author[Sara A. Webb $\&$ Simon R. Goode]   
{Sara A. Webb$^1$, Simon R. Goode$^1$$^2$ }

\affiliation{$^1$ Centre for Astrophysics and Supercomputing, Swinburne University of Technology, Hawthorn, Melbourne, Australia. \\ $^2$ OzGrav ARC Centre of Excellence for Gravitational Wave discovery, Swinburne University of Technology, Hawthorn, Melbourne 3122, Australia \\ email: {\tt swebb@swin.edu.au}}

\pubyear{2015}
\jname{Astronomy in Focus, Volume 1} 
\editors{}
\begin{document}

\maketitle

\begin{abstract}
With the volume and availability of astronomical data growing rapidly, astronomers will soon rely on the use of machine learning algorithms in their daily work. This proceeding aims to give an overview of what machine learning is and delve into the many different types of learning algorithms and examine two astronomical use cases. Machine learning has opened a world of possibilities for us astronomers working with large amounts of data, however if not careful, users can trip into common pitfalls. Here we'll focus on solving problems related to time-series light curve data and optical imaging data mainly from the Deeper, Wider, Faster Program (DWF). Alongside the written examples, online notebooks will be provided to demonstrate these different techniques. This guide aims to help you build a small toolkit of knowledge and tools to take back with you for use on your own future machine learning projects.
\end{abstract}

\firstsection 
\section{Introduction}

In the field of artificial intelligence, machine learning focuses on using data and algorithms to mimic the way humans would typically learn, improving accuracy over time. Via machine learning, we can automate analytical models, taking advantage of algorithmic ability to learn from data and identify patterns with minimal human input. \\  

Machine learning has already begun to be adopted by a wide range of sub-disciplines in astronomy, and is well established in some areas including it's use in transient astronomy as outlined in the advanced review by \cite{fluke2020}. Note this work is not a comprehensive review of all techniques, rather a compilation of specific examples used within established transient astronomy programs. \\

In the era of large current and upcoming time-domain surveys, the classification and discovery of transient sources will rely on machine classification to handle large amounts of collected data. Current ground-based surveys such as the ZTF, DES and ASAS-SN scan thousands of square degrees per night, amounting to petabytes of data annually. Recently the Panoramic Survey Telescope and Rapid Response System Survey (Pan-STARRS) delivered the first-petabyte scale optical data release (\cite{Bellm2018, DES2016, Shappee2014, Stubbs_2010, Chambers2016}). \\

Space-based time-domain missions have provided unprecedented volumes of photometry, light curves, and proper motions for Galactic sources, with \textit{Kepler} and K2 targeting $\sim$400,000+ individual stars, Transiting Exoplanet Survey Satellite 
(TESS, \cite{Ricker2009})
is expected to target at least 200,000 of the $\sim$9.5 million catalogue sources. The space-based mission, \textit{Gaia}, is already releasing almost 2 billion sources \cite{Borucki2010, Howell2014, Stassun2018}. \\

Overcoming the mining challenges of these increasing amounts of data to not only identify and catalogue the multitude of known transient types but to discover additional new or anomalous sources is paramount to the success of future large transient surveys and time-domain science. This will become especially important with the upcoming Vera Rubin Observatory Legacy Survey of Space and Time (LSST, \cite{LSST2009}).
LSST is a planned 10-year survey, imaging the entire Southern sky every three nights. LSST will generate millions of alerts each night, with billions of light curves continually updated or created.  \\

\begin{figure}[b]
    \centering
    \includegraphics[width=15cm]{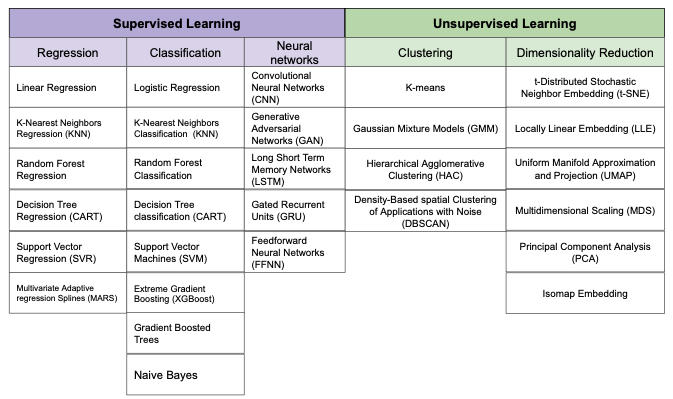}
    \caption{Representation of supervised verses unsupervised learning and the specific examples of each. }
    \label{fig:ML_types}
\end{figure}

Machine learning can be broken down broadly into either supervised or unsupervised algorithms, each pertaining subcategories beneath them. See Figure \ref{fig:ML_types} for more details on the specific types of categories which will be explored below in more detail relating to the two real-data examples later on in the paper. For a comprehensive overview of techniques and uses in astronomy see \cite{Kembhavi2022}, \cite{Alzubi2018}, and \cite{Mathew2020}. \\

\section{Overview}
\subsection{Supervised Machine Learning}

One heavily used subcategory of machine learning in astronomy is supervised machine learning. This uses labelled data sets to train algorithms to classify data or predict outcomes. Traditionally a supervised algorithm would tune itself around the labelled data sets generating a function to map new inputs to likely outputs. \\

Supervised machine learning can be separated into three main branches: 1) Regression algorithms, 2) classification algorithms and 3) Neural networks. It is important to note that neural networks can be used to solved both regression and classification problems.\\

Regression algorithms focus around establishing the relationship between a single dependent variable that is dependent on several independent ones. Both linear and non-linear regression can be used for supervised  learning. An example of a commonly used non-linear regression algorithm is that of Decision Trees (\cite{Quinlan86}). 
Decision Trees work by breaking down data into either decision or leaf nodes. Decision nodes are where the sub-node splits into further sub nodes, whereas the leaf nodes represent a final decision or terminal node. Once a decision tree is created on features from labeled data, the algorithm is able to assign a predicted value or outcome on subsequent data (\cite{Quinlan86}). \\

Classification algorithms are able to learn from a given labeled dataset to sort, assign and classify new data into a specific given number of classes or groups. Unlike regression algorithms which output continuous values, classification algorithms will only result in an assigned class or group. Classifications can be either binary or multi-class and are dependent on the problem being approached. A common classifier used in astrophysical contexts, including transient astronomy, is via a  Decision Tree. Decision trees work by creating internal nodes and leaf nodes of grouping data around the features of the data. The internal nodes represent the conditions (e.g. certain features present or not) and the leaf nodes represent the decision made based on the conditions. These algorithms are useful for constructing easy to interpret tree like models around known data which can be used to classify new data. \\

As data is often plentiful in astronomy, supervised algorithms are ideal for completing the `hard lifting' in classifying and sorting astronomical data sets. \\

One area which as heavily utilizes supervised algorithms is in the identification of variable sources. Variable stars and quasi-stellar objects have been identified from light curves via multivariate Gaussian mixture models, random forest classifiers, support vector machines, or Bayesian neural networks (\cite{Debosscher2007,richards2011,Kim_2011, Pichara2012,Bloom_2012,Pichara2013, Kim2016, Mackenzie2016, Muthukrishna2022}). \\

All of the aforementioned work successfully uses classification of objects via supervised algorithms, which were trained on light curve extracted features. Features represent a set of measurable properties/characteristics of the light curves being studied. The most common features used in earlier works are available within the python package `\textit{FATS}' by \cite{Nun2015}. \\

Classification of non-folded light curves of extragalactic transient sources has also been explored, moving away from selecting the class of the object by fitting analytical templates built from a set of known sources \cite{richards2011, Karpenka2012, Lochner2016, Narayan_2018, moller2016}. While these techniques work well for catalogues of light curves, they cannot easily be applied to real-time data. Real-time classification of supernovae by \cite{Muthukrishna2019} and \cite{moller2019} has shown the effectiveness of deep recurrent neural networks, without the need to rely on extracting computationally expensive features of the input data. \\

Another heavily explored use of supervised learning is for the determination or `real' or `bogus' sources in transient astronomy. These algorithms work by inspecting an image, or features of an image, and decide whether or not the image is of a genuine astrophysical source. \\
 \\

\subsection{Unsupervised Machine Learning}

With unsupervised machine learning, algorithms learn patterns from unlabeled data. These unsupervised algorithms self organise and create groupings or classes based on patterns exhibited as neuronal predilection or probability densities. These techniques are instrumental in finding like among like within a large data set. \\

Unsupervised machine learning can be separated into two main branches: 1) Clustering algorithms and 2) Dimensionality reduction algorithms. \\

Clustering algorithms work by grouping data into like clusters using features extracted from the data. The ultimate goal of clustering is to group like data together, and to isolate different clusters present within data. There are four main types of methods in clustering unlabelled data. The first is via density-based clustering, which works by identifying areas in feature space of high concentrations of data points. Distribution-based clustering takes the approach that all data points are part of the expected number of clusters, and works by calculating the probability that they belong to any given cluster. This works by using distance between points in feature space to determine the locations of clusters. Centroid-based clustering works by isolating the likely centroids within the data in feature space and determining relation to each cluster via distance metrics. Finally, hierarchical-based clustering works by organising data and groupings with a top--down approach, to insure groupings of varying densities are still identified as separate clusters. \\

In astronomy, clustering techniques have been used within recent large data sets to isolate known transient and variable sources using light curves (\cite{Valenzuela2018,Giles2019,Galarza2020}). 
This ability to cluster similar data will help identify previously unknown variables and transient events in the era of large astronomical surveys. As such, it will be invaluable to meaningfully and quickly quantify the expected large volume of short timescale events to help assist in follow-up priority assignment (\cite{LSST2009}).\\

Both have been explored within astronomical data and proved successful in providing meaningful insight into large data sets. \\

\section{Examples of Applications in Astronomy }

Here we outline and present resources to familiarize yourself with applying supervised and unsupervised machine learning to an observational astronomy data set. For both examples we'll be using different aspects of the Deeper Wider Faster (DWF) programs data. \\

The DWF program was developed to explore the fast dynamic universe, through multi-wavelength, multi-facility, real-time observations. The program is designed to run over $\sim$1 week observation blocks, at least once a year. The program is optimised to detect fast transients in real-time, and provide rapid follow-up with additional facilities. This first application of the programs brings up to the use of supervised machine learning in near-real time. \\

\subsection{Supervised Learning: Removal Of BOgus (ROBOT pipeline) for the Deeper Wider Faster program} 

This section we outline one example of supervised learning from \cite{Goode2022}, which can be further explored using the interactive notebooks on GitHub\footnote{\href{https://github.com/simongoode/ROBOT-pipeline}{https://github.com/simongoode/ROBOT-pipeline}}. \\

During a typical DWF run, raw optical data (primarily from the CTIO Dark Energy Camera) is transferred in near real-time to Swinburne's OzSTAR supercomputer (\cite{vohl2017}). Each DECam image is comprised of 60 individual ccds, each 4K $\times$ 2K pixel resolution. The footprint of the imaging is large, covering $\sim$ 3 squared degrees. \\

\begin{figure}
    \centering
    \includegraphics{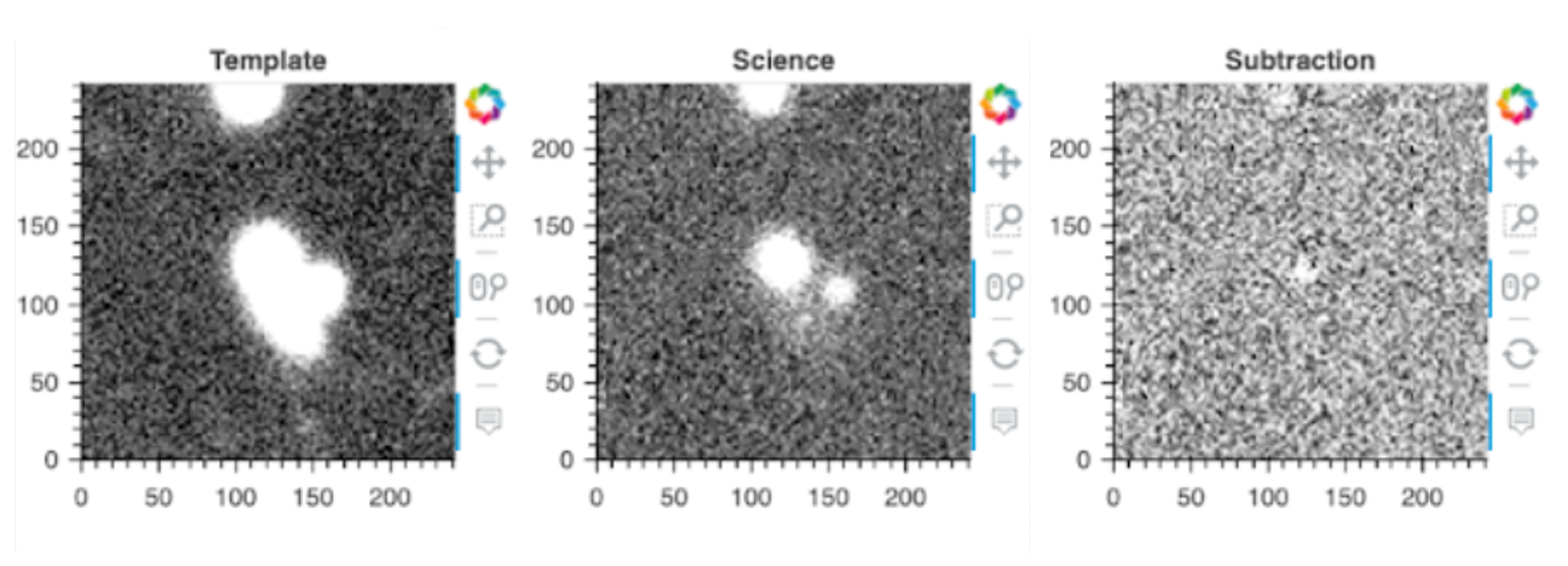}
    \caption{Example of real time processed candidate images, taken from a past operational run of DWF. Each panel is a small, 121  $\times$ 121 pixel image that corresponds to $\sim$ 30 $\times$ 30 arcsec on the sky centred on the candidate. The left panel is the (deeper) template image taken at a time previous to the DWF observations, the central panel is the current science image of the sky taken minutes earlier, and the subtraction image is the digital subtraction of the two images. All constant flux is subtracted, and any flux difference, e.g. from a transient source, will remain in the subtracted image.}
    \label{fig:TSS_dwf}
\end{figure}

Once the data has arrived on OzSTAR, it is processed for calibrations and made `science ready' before being ingested through the \textit{Mary} pipeline (\cite{andreoni2017}). \textit{Mary} performs alignment and difference imaging between template and science images, and rapidly identifies transient candidates from positive residuals in the subtractions. During a single DWF observation run, hundreds of thousands of transient candidates are flagged through difference imaging, and it is crucial that promising candidates are inspected manually before triggering space and ground-based telescopes for follow-up observations. A large problem encountered during early DWF runs was the the immense data volume, which often exceeded the capability to all be accessed manually. With hundreds of thousands of candidate found in the processing, human inspectors were faced with more data then physically possible to evaluate without assistance. Astronomers were given several key pieces of information in a DWF run, one of which is the `postage stamp' images of candidate sky location. Figure \ref{fig:TSS_dwf} shows an example of a transient candidate as processed in real time during a DWF run. Each of the three `cutouts' is important for determining the realness of the the source, and what type of source it likely is. \\

The Removal Of BOgus Transients (ROBOT) pipeline was developed to significantly reduce the number of candidates needing human inspection, and rapidly improve the efficiency of candidate inspection during DWF observational runs (\cite{Goode2022}). The ROBOT pipeline aimed to work as an intermediary step between the processing/candidate identification and the image inspection preformed by astronomers. Due to the nature of the work, the uncertainties in sky conditions, and tendencies towards compression artifacts, a large majority of the data is often deemed as `Bogus' or False positive alerts. ROBOT works to identify the astrophysical realness of objects, and filter only those of the highest likelihood through to the human inspector. \\

A deep Convolutional Neutral Network (CNN) was chosen to tackle this task, as historically CNNs have proven to be excellent for 2 dimensional data structures, like image data, and have a proven track record of reliable uses in image classification. The first step in building the ROBOT frame work was compiling and labeling large amounts of past DWF data into either real or bogus categories, and even specific source types. \\

Labelling of data occurred over several sessions by multiple expert astronomers. Using their past knowledge and experience, specifically around the DWF program, ensured contextually informed decisions. A total of 2952 candidate images containing template, science subtraction images were used. Out of the initial samples it was found 2250 were labelled unanimously by experts as bogus, and only 702 candidates were labelled as real. To limit the bias in the eventual network, we chose to balance the labelled data by using data augmentation. To do this, both the real and bogus images where included multiple times in the data set, but in different random augmentations including rotating, mirroring and  translation. Each labelled set contained 5000 samples. \\

\begin{figure}
    \centering
    \includegraphics{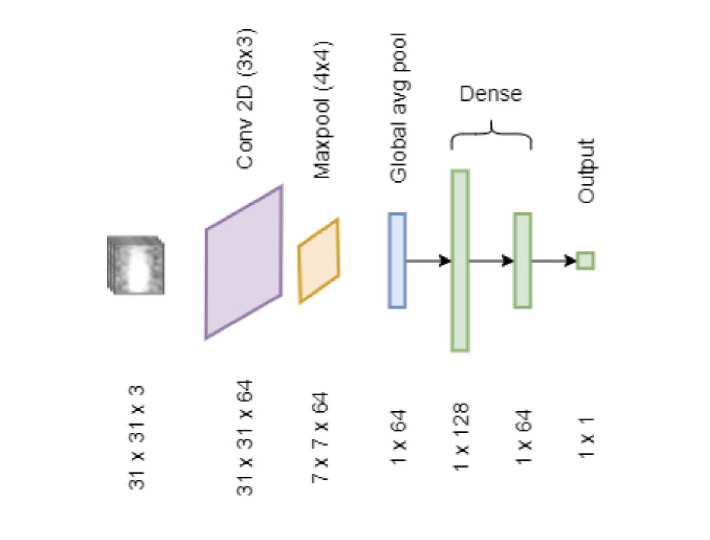}
    \caption{Figure from \cite{Goode2022}. Architecture diagram of the highest performing model found during testing. The model takes in cropped triplet images (31 $\times$ 31 $\times$ 3) as input into a single 2D convolution layer, with 64 filters (3 $\times$ 3). A 4 $\times$ 4 maxpool layer passes 7 $\times$ 7 $\times$ 64 information to a global average pooling layer, which functions as both a dropout and flatten layer. This pooling layer passes the information to two fully connected dense layers, before providing a single output; the probability that the input triplet images belong to a real object.}
    \label{fig:robot}
\end{figure}

Using the labelled data, \cite{Goode2022} 
trained an initial 60 different CNN model architectures, each which slightly different combinations of layers, convolutions and hyperparameters. Each of the architectures was evaluated on their initial performance using the Matthews Correlation Coefficient, which took into account false positives, false negatives, true positives and true negatives to find a the architecture which preformed the best. It was found the best model was a `1c$\_$2d' architecture, which consisted of 1 convolution, and 2 dense layers which are fully connected. The final architecture can be seen in Figure \ref{fig:robot}. The final algorithm preforms regression, returning an overall score between 0 and 1 for each the candidates passed through. The scores can then be used to determine the likelihood of it being either real or bogus, and can be used a classifier by setting the limits at which a final label of real or bogus is assigned to each. 

By implementing ROBOT into DWF operations, the total time needed to inspect candidates was dramatically reduced, speeding up the human hours needed to make meaningful discoveries in real-time. Astronomy is uniquely positioned with the vast amounts of archival data able to be used in creating methods for automated discovery.

\newpage
\subsection{Unsupervised Learning: Anomaly detection in lightcurves for the Deeper Wider Faster program} 

This section we outline one example of unsupervised learning from \cite{Webb2020}, which can be further explored using the interactive notebooks on GitHub\footnote{\href{https://github.com/sarawebb/ML\_lightcurve\_clustering}{https://github.com/sarawebb/ML\_lightcurve\_clustering}}. \\

Although DWF is focused on chasing the fastest transient in near real-time during the observational runs, the data is still fully processed and explored systematically for other science objectives. One part of the post run processing is the production of light curves using the optical imaging, for every source detected, not just new transient sources. For a standard field, upwards of 100,000+ sources are present. To meaningfully evaluate this volume of sources, analytic and automated algorithms to identify sources of interest. \\

One exciting aspect of the DWF optical data is the cadence at which it is collected. Using continuous 20 second exposures, the lightcurves generated from this data have an average time of $\sim$ 60--75s between data points. This cadence allows high time resolution of transient and variable events. One area of great interest is identifying new or under explored transients and variable sources in the unique DWF data. To explored possibly unknown events we needed to design a flexible algorithm which could identify lightcurves which were anomalous to the majority. \\

We choose to use Hierarchical Density- Based Spatial Clustering of Applications with Noise (HDBSCAN, \cite{McInnes2017}).The theoretical method behind this algorithm was first proposed by \cite{Campello2013}. HDBSCAN takes the approach of Density-Based Spatial Clustering of Applications with Noise (DBSCAN) and converts it into a hierarchical clustering algorithm by varying the value of epsilon ($\epsilon$) to identify clusters of varying densities. \\

The power of HDBSCAN lies within it's ability to identify clusters of varying densities within a dataset. This is a valuable tool when working with diverse data such as lightcurves. Although we expect the majority of the sources to be unchanging over the short observational time periods, those which do change will do so in a variety of different ways, and not have consistent percentages of the data they occupy. Another advantage of HDBSCAN is it's ability to identify data which is highly anomalous and not belonging to any of the identified clusters. Before we are able to cluster the DWF lightcurves, we first needed to identify what features we wanted to use to describe each. \\

\begin{figure}
    \centering
    \includegraphics{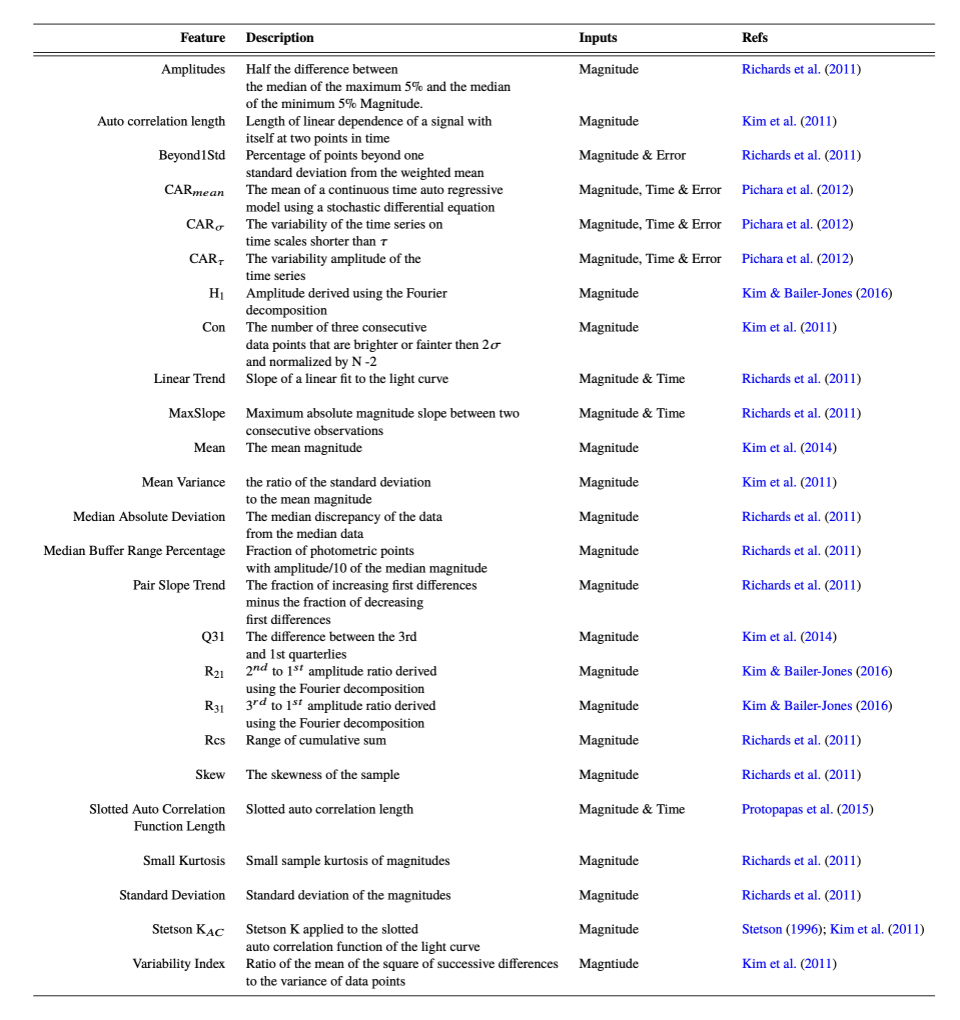}
    \caption{ Features used in \cite{Webb2020} and the properties of the light curves they represent.}
    \label{fig:features}
\end{figure}

Features represent a set of measurable properties/characteristics of the light curves being studied. In this work we extracted a uniform set of features across all light curves for two purposes 1) reduce the dimensionality of the light curves and 2) allow for direct comparison between light curves that may be on different time scales with different sampling properties. We chose to  use a mixture of normalised features developed and used primarily for the identification of variable stars and quasi-stellar objects, aiming to cluster variable and periodic sources, and have highly anomalous sources encapsulated in the unclusterable `noise' unidentified by HDBSCAN. We began the feature selection by working on a sample sub sample of data, and calculating multiple different previously used, and data specific features. Using principle component analysis, we selected the top 25 with the largest eigenvalues. The chosen features are shown in Table \ref{fig:features} and explained in detail in \cite{Webb2020}. We extracted these 25 unique features from each light curve using mostly using \textit{FATS}/\textit{FEETS} packages and some in-house routines (\cite{Nun2015}). \\

Using the lightcurves features we tested multiple configurations of HDBSCAN, changing minimum cluster size and distance metric type. After the preliminary tests we decided on the use of a minimum cluster size of 5 and the use of the Euclidean distance metric, for its intrinsic ability to calculate the shortest distance between points. These were chosen in an effort to create as many distinct clusters in our feature space as the algorithm will allow to limit the outliers to very low density regions.\\

In \cite{Webb2020} two separate fields/run types were explored 1) the DWF ‘J04-55 field' which was data collected using a staring method on the telescope, and 2) `Antlia field' collected using interleaved dithered. Both observational methods have their merits and uses, and we wanted to confirm that the clustering methods would work to identify astrophysical variability as well as variability caused by observational effects such as dithering. \\

\begin{figure*}
    \centering
    \includegraphics[width=16.5cm]{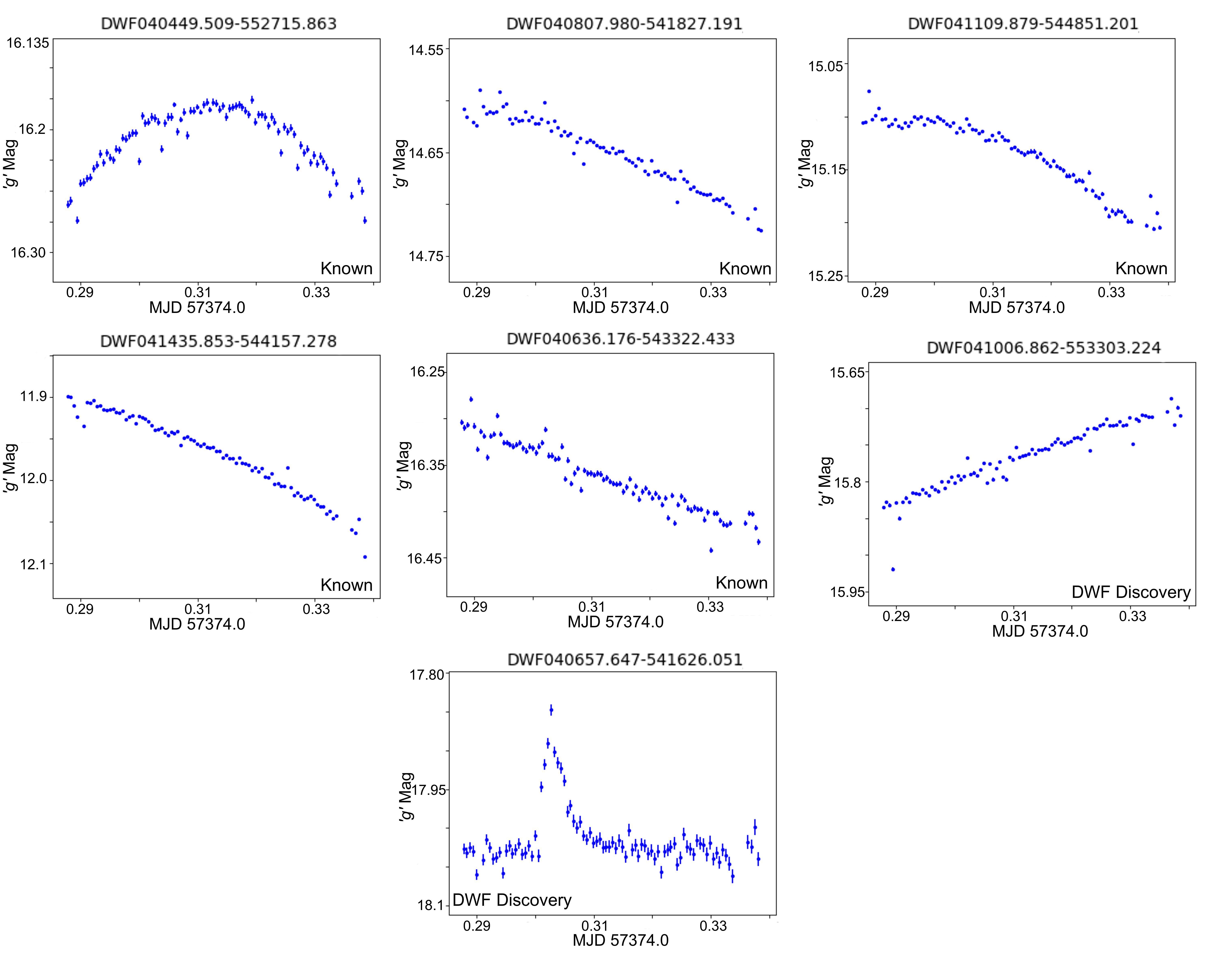}
    \caption{Four previously known and two newly discovered variable/transient sources present in the unclustered noise within the J04-55 field analysis.
    }
    \label{fig:subJ04-55noiseplots}
\end{figure*}

The clustering methods via HDBSCAN proved extremely successful in identifying not only distinct groupings of astrophysical sources, but also clustering lightcurves which were affected by observational effects such as dithering, blended sources or cosmic rays. Table \ref{tab:J04-55cluster} breaks down the cluster types identified using the DWF ‘J04-55 field' of 23,199 lightcurves, with the distant clusters containing sources of unchanging magnitudes, or those at detection thresholds or CCDs edges. Interestingly in this field, the true astrophysical variable and transient sources were unable to be clustered and identified as noise.  Figure \ref{fig:subJ04-55noiseplots} shows 7 such sources extracted from the grouping of noise. 

For full analysis and the results from the `Antlia field' , including the use of \textit{Astronomaly} and t-SNE's, and results see \cite{Webb2020}.

\section{Conclusions}

Machine learning has already proven to be extremely powerful in it's ability to assist astronomers in discovery, and will only continue it's growth into more astronomical use cases. It is always important to note that machine learning isn't always a one solution fits all. It should be considered and applied with a great deal of care, to insure the problem tackled is solved in an efficient and unbiased manner. For those just beginning to explore the use of artificial intelligence in astronomical work, we highly recommend the use of existing frame works to evaluate the effectiveness of different methods. For the use of anomaly detection, the \textit{Astronomaly} package is a flexible framework for use on both imaging and light curve data (\cite{Lochner2021}). It is undeniable that machine learning will shape the future of astronomy, with several large surveys already relying on intelligent algorithms.

 \begin{table}
    \centering
    \begin{tabular}{lccr}
    \hline 
     \textbf{Description of} & \textbf{Cluster} &  \textbf{$\#$ of} & \textbf{$\%$ of}   \\
     \textbf{light curves} &  \textbf{ID} & \textbf{Light Curves} & \textbf{Sources} \\
    \hline 
    \hline 
    Faint sources &   \\
     at detection  & Cluster 0  &   8   & 0.03$\%$    \\
     threshold \\
    Sources near   \\
    CCD edge &  Cluster 1  &   144   & 0.62$\%$    \\
    Steady light curves & Cluster 2& 22909 & $>$98.7$\%$ \\
    Real and &  \\
    photometrically  & Unclustered   &   138   & $<$ 0.59$\%$       \\
    affected light curves \\
    \hline
    \end{tabular}
    \caption{The details of each of the three clusters identified by the HDBSCAN algorithm.  The description of the light curves refers to both the light curve and information gathered from individual cutouts of the detection images.  unclustered represents light curves unable to be identified to a cluster. }
    \label{tab:J04-55cluster}
\end{table}

\end{document}